\documentclass[acmsmall,natbib=false,nonacm]{acmart}

\usepackage{listings}
\usepackage{xcolor}

\usepackage{xcolor}
\definecolor{lbg}{RGB}{240,240,240}
\usepackage{tcolorbox}
\usepackage{here}
\usepackage{float}
\usepackage{array}

\AtBeginDocument{%
  }

\renewcommand\footnotetextcopyrightpermission[1]{}
\setcopyright{none}

\acmConference{}{}{}
\acmBooktitle{}
\acmYear{}
\acmMonth{}
\acmVolume{}
\acmNumber{}
\acmArticle{}
\acmPrice{}

\makeatletter
\def\@copyrightspace{\relax}
\makeatother

\RequirePackage[
  datamodel=acmdatamodel,
  style=acmauthoryear,
  ]{biblatex}

\begin{document}

\title{QuCheck: A Property-based Testing Framework for Quantum Programs in Qiskit}

\author{Gabriel Pontolillo}
\orcid{0000-0002-4529-2903}
\affiliation{%
  \institution{Department of Informatics, King's College London}
  \streetaddress{Strand}
  \city{London}
  \country{United Kingdom} 
  \postcode{WC2R 2LS}
}
\email{gabriel.pontolillo@kcl.ac.uk}

\author{Mohammad Reza Mousavi}
\orcid{0000-0002-4869-6794}
\affiliation{%
  \institution{Department of Informatics, King's College London}
  \streetaddress{Strand}
  \city{London}
  \country{United Kingdom} 
  \postcode{WC2R 2LS}
}
\email{mohammad.mousavi@kcl.ac.uk}

\author{Marek Grzesiuk}
\orcid{0009-0003-7296-3113}
\affiliation{%
  \institution{Department of Informatics, King's College London}
  \streetaddress{Strand}
  \city{London}
  \country{United Kingdom} 
  \postcode{WC2R 2LS}
}
\email{marekgrzesiuk99@gmail.com}

\begin{abstract}

Property-based testing has been previously proposed for quantum programs in Q\# with QSharpCheck; however, this implementation was limited in functionality, lacked extensibility, and was evaluated on a narrow range of programs using a single property. To address these limitations, we propose QuCheck, an enhanced property-based testing framework in Qiskit. By leveraging Qiskit and the broader Python ecosystem, QuCheck facilitates property construction, introduces flexible input generators and assertions, and supports expressive preconditions. We assessed its effectiveness through mutation analysis on five quantum programs (2-10 qubits), varying the number of properties, inputs, and measurement shots to assess their impact on fault detection and demonstrate the effectiveness of property-based testing across a range of conditions. Results show a strong positive correlation between the mutation score (a measure of fault detection) and number of properties evaluated, with a moderate negative correlation between the false positive rate and number of measurement shots. Among the most thorough test configurations, those evaluating three properties achieved a mean mutation score ranging from 0.90 to 0.92 across all five algorithms, with the false positive rate between 0 and 0.04. QuCheck identified 36.0\% more faults than QSharpCheck, with execution time reduced by 81.1\%, despite one false positive. These findings underscore the viability of property-based testing for verifying quantum systems.

\end{abstract}




\maketitle

\section{Introduction}

\subsection{Background and Motivation}

Testing and debugging are laborious and costly \cite{DBLP:journals/ibmsj/HailpernS02,Kafle2014}, yet indispensable \cite{NIST2002}, part of software development. Designing tests is challenging for classical systems. This challenge intensifies with quantum systems, where capturing the underlying principles and mapping the intricate input-output relationships are particularly difficult.

Property-based testing \cite{Hughes2000} is an approach to mitigate this problem; it has been successfully applied to various types of classical systems, including telecommunication software, vehicular communication stacks, and databases \cite{experiencesquickcheck}.  Recently, an initial prototype called QSharpCheck \cite{DBLP:conf/icse/Honarvar0N20} was developed for the Microsoft Q\# language \cite{QsSpec2020} to extend property-based testing to quantum systems. QSharpCheck has attracted attention and been utilised in subsequent research \cite{Castro2021, Hub2024}. However, this initial prototype had limited functionality: assertions were restricted to comparisons of single qubits on the computational basis, it lacked support for generating multi-qubit states or oracle circuits, and precondition specifications were tied to input generation.

\subsection{Contributions}

In this paper, we present QuCheck, a redesigned property-based testing framework for the popular Qiskit platform \cite{qiskit2024}. It addresses the shortcomings of QSharpCheck and makes it available to the larger community of researchers and practitioners using Qiskit. Moreover, we design and carry out experiments to evaluate and compare the efficiency and effectiveness of our newly developed framework with its past incarnation, QSharpCheck. To summarise, our contributions are listed below: 

\begin{enumerate}
    \item QuCheck, an improved property-based testing framework for for Qiskit, extending previous property-based testing approaches for quantum programs through the introduction of statistical corrections, optimisation via circuit deduplication and greedy measurement insertion, support for variable size circuits, diverse and customisable input generators with the ability to generate input oracles, and assertions that compare multiple qubits and bases;
    \item Evaluation of the effectiveness of property-based testing for quantum programs;
    \item Analysis on how the number of properties, inputs and measurement shots relates to the effectiveness of property-based testing in the quantum computing context; and 
    \item A qualitative and quantitative comparison of QuCheck with QSharpCheck.
\end{enumerate}

\subsection{Research Questions and Results}
\subsubsection{Research Questions} We pose the following research questions and answer them through carefully designed experiments to evaluate the effectiveness of property-based testing for quantum computing in general, and of QuCheck in particular, by comparing it with QSharpCheck.    
\begin{itemize}
    
    \item \textbf{RQ1:} How does the \textit{thoroughness} of a property-based testing suite affect the ability to identify faults in quantum programs?
    
        \begin{itemize}
        
            \item \textbf{RQ1.1:} How does the \textit{number of properties per program} affect the ability to identify faults in the quantum programs?
        
            \item \textbf{RQ1.2:} How does the \textit{number of inputs per property} affect the ability to identify faults in quantum programs?
        
            \item \textbf{RQ1.3:} How does the \textit{number of measurement shots} affect the ability to identify faults in quantum programs?
            
        \end{itemize}

    \item \textbf{RQ2:} Is property-based testing \textit{effective} for the identification of faults in quantum programs?
    
    \item \textbf{RQ3:} How does QuCheck compare to QSharpCheck?
    
        \begin{itemize}
        
            \item \textbf{RQ3.1:} What are the qualitative differences between QuCheck and QSharpCheck?
        
            \item \textbf{RQ3.2:} How does QuCheck compare to QSharpCheck in identification of faults in quantum programs?
             
        \end{itemize}

\end{itemize}

To determine property-based testing's \textit{effectiveness} at identifying quantum program faults, we employed mutation score and execution time as our primary metrics. Additionally, the \textit{thoroughness} of the suite is determined by the number of properties, inputs per property, and number of measurement shots, which we investigate through three sub-questions, labeled RQ1.1 to RQ1.3.

\subsubsection{Results} Our experiments revealed that increasing the thoroughness of the test suite increases the ability to identify faults while reducing the false positive rate. Increasing the number of properties evaluated showed a strong positive correlation with fault detection ($r$ = 0.532). Similarly, increasing the number of inputs showed a moderate positive correlation with fault detection ($r$ = 0.239). While increasing the number of measurement shots only weakly improved fault detection ($r$ = 0.121), it was effective in reducing false positive rates ($r$ = -0.266). 

Overall, property-based testing proved to be an effective method for identifying faults in quantum programs, as evidenced by high mutation scores observed across the most thorough configurations (0.90 to 0.92), accompanied by a low false positive rate (0 to 0.04) and feasible execution times.

QuCheck provides a more expressive and extensible testing framework than QSharpCheck that is able to detect more mutants, with a reduced execution time. This is supported by its ability to enable more customisable input generation, precondition checking, expressive assertions, and the execution of complex operations, such as those found in metamorphic properties. As a result, QuCheck identified 36.0\% more faults than QSharpCheck with an 89.0\% reduction in average execution time, though it did produce one false positive.   

The lab package for QuCheck can be accessed at \cite{pontolillo_grzesiuk_mousavi_2024}.

\subsection{Paper Structure}

The remainder of this paper is organised as follows. Section \ref{related_work} reviews related work on property-based testing in both classical and quantum contexts, contrasting them to other testing methodologies proposed for quantum programs. In Section \ref{property-based-testing-for-quantum-programs-secion}, we describe property-based testing in more detail, outlining the advantages of its application in the context of quantum computing. In Section \ref{QuCheck_section}, we describe the property-based testing framework that we have developed, detailing its overall architecture, property specification, test execution process, and statistical analysis. In Section \ref{experiment-design}, we outline the experimental setup for each research question, including the quantum programs used to evaluate our testing framework. We present and analyse the experimental results in Section \ref{results}. Section \ref{threats} describes the threats to the validity of our work. Finally, in Section \ref{conclusion} the paper concludes with a brief discussion of the results, the impact of the work, its limitations and future directions.  

\section{Related Work}\label{related_work}

There has been considerable progress in the field of testing for quantum programs, with multiple approaches proposed. Nevertheless, the fundamental challenges posed by quantum mechanics remain to be addressed. In this section, we review related work in three areas: classical property-based testing, property-based testing for quantum programs, and general testing approaches and tools in the quantum domain. 

\subsection{Property-based testing for classical programs}

Property-based testing was initially introduced in the Haskell programming language through a tool named QuickCheck \cite{Hughes2000}, enabling the automated testing of program characteristics and invariants via the generation and execution of random test cases. Over time, this methodology has been ported across multiple programming languages, including Scala \cite{scalacheck}, Prolog \cite{prologcheck}, and Erlang \cite{quickcheck2}. It has also been adopted commercially, with applications in software testing across the telecommunications \cite{telecoms}, e-commerce, and automotive industries \cite{experiencesquickcheck}.

\subsection{Property-based testing for quantum programs}

Honarvar, Mousavi, and Nagarajan \cite{DBLP:conf/icse/Honarvar0N20} proposed QSharpCheck, a property-based testing framework for Q\#, allowing the specification of properties in a plain text file, which are parsed and executed by their tool, yielding a counterexample when a property does not hold. QSharpCheck’s approach to specifying property inputs and preconditions is limited to defining the range of $\theta$ and $\phi$ for each individual qubit. This approach does not allow for the generation of entangled states, or more general inputs that may be required, such as oracle circuits. Furthermore, the statistical analysis performed does not apply a correction (such as the Holm-Bonferroni correction) for the multiple test problem to control for the family-wise error rate when performing sequential statistical tests. 

In our previous work \cite{PontolilloM22}, we curated a set of benchmark quantum programs and compared the performance and popularity of Cirq, Qiskit, and Q\#:
\begin{itemize}
	\item Popularity was investigated by analysing quantum computing repositories on GitHub, finding Qiskit to be more prevalent in the top hundred results across all the tested sorting options. 
	\item We identified a disparity in performance when running the algorithms between the different platforms, finding Q\# to be least performant, citing our experimental results for the execution of the quantum phase estimation algorithm. 
\end{itemize}

In \cite{deltadebuggingproperty}, we investigated the application of property-based testing and delta-debugging for quantum programs, where a set of changes between a passing and failing quantum circuit are recursively tested to isolate a minimal set of changes that cause a failure. We applied a property-based test oracle, comprised of a set of three properties of the circuit under test to determine whether a tested subset of changes passes or fails. The oracles' ability to differentiate between correct and faulty circuits was demonstrated through its successful integration with delta debugging. However, the tests were implemented on an ad-hoc basis, and would have benefited from a rigorous property-based testing framework.  

The findings mentioned above, the gaps identified in prior work, and the accessibility of Python motivated our decision to develop a property-based testing framework for quantum circuits in Qiskit. To the best of our knowledge, no other published property-based testing approaches currently exist for Qiskit.

\subsection{Other testing techniques for quantum programs}

Researchers have proposed various approaches for testing quantum programs, including:

\begin{itemize}
    
\item Search-based testing \cite{qusbt, muttg}, which uses genetic, or evolutionary algorithms to generate test cases. The goal of these approaches differs from that of QuCheck, where the focus \textit{is on efficiency}. Using a program specification that describes the input-output relation of the programs: MutTG maximises the number of mutants killed with a fixed number of test cases, and QuSBT attempts to generate a minimal test suite that can kill all mutants from a set. In contrast, QuCheck \textit{verifies the correctness} of a program using a set of invariants and logical rules of a quantum program encoded into properties.  

\item Fuzz testing \cite{wang2018quanfuzzfuzztestingquantum}, which analyses the source code to identify branches associated with measurements in the circuit, then dynamically executes the circuit, receiving feedback to generate test inputs with higher branch coverage than random testing. In contrast, property-based testing explores a wide range of inputs to verify specified program properties, rather than maximising for branch coverage directly, though with sufficient inputs, it will eventually traverse all branches. 

\item Combinatorial testing \cite{qucat} generates computational basis test suites of a specified strength, or until a failure is found. While combinatorial testing is effective for evaluating discrete inputs; QuCAT, unlike QuCheck, lacks the ability to generate continuous input states or states in alternative bases. 

\end{itemize}

Static analysis techniques have been developed for quantum circuits \cite{qchecker, lintq}. These offer a resource-efficient way to detect faults before circuit execution, helping to address scalability challenges in quantum software development. However, certain faults will only manifest at run time, after circuit initialisation. In such cases, our approach, which executes the circuits, is well suited to detect these faults. 

QuraTest \cite{quratest} is an automated test case generation tool that leverages three test case generators, UCNOT, IQFT, and random, in order to generate inputs that are entangled, or have a phase. These inputs are evaluated based on input diversity, output coverage, and mutation score. On the other hand, QuCheck is a property-based testing framework that facilitates the implementation and verification of properties, through the provision of various input generators, ability to check preconditions on inputs, and statistical assertions. For the purposes of our experiments, we use a simulator and initialise qubits to state vectors provided by the random statevector generator. However, our framework also includes a replication of QuraTest's UCNOT input generator, that initialises a random state by applying U and CNOT gates, to provide compatibility for when working on real hardware.

Metamorphic testing \cite{metamorphic} has been proposed for oracle quantum programs, they define metamorphic relations using the properties of the program under test, demonstrate how to encode them into a quantum circuit. In contrast, while we implement some metamorphic properties for our case studies, they are not encoded into the quantum circuit. Instead, two executions of the circuit are performed with different inputs, and the outputs are compared.

Proq \cite{proq} is a runtime assertion scheme that employs projection-based assertions to verify quantum states throughout program execution. Proq's assertions can verify quantum states more efficiently compared to statistical assertions, but they are less suitable for property-based testing, as each randomly generated input requires a different projection based assertion (if it modifies the asserted subspace). However, this approach may be implemented for the verification of invariants, where the subspace is not modified by the input, thus allowing for a fixed projection-based assertion. In contrast, our approach incorporates statistical assertions in the spirit of Huang and Martonosi \cite{Martonosi2019statisticalassertions}, but employs local measurements conceptually related to those described by Yu \cite{yu2020quantumclosenesstestingstreaming}, which are more efficient to verify, but are not able to differentiate all multiple qubit states.

Mutation testing tools such as Muskit \cite{muskit} and QMutPy \cite{qmutpy} have been previously proposed and applied for the evaluation of test suites \cite{DBLP:conf/icse/Honarvar0N20, multiSubroutine, quratest}, they apply mutation operators that modify the underlying quantum circuits of a program, such as through the addition, removal or replacement of quantum gates at random locations in the circuit. In our experiments, we chose to use QMutPy due to its compatibility with the structure of our case study code. Specifically, QMutPy allows mutations to be applied to the code that generates quantum circuits, rather than directly to the circuits themselves.

To the best of our knowledge, none of the discussed methodologies consider oracles, sub-circuits, or unitary gates \textit{as potential inputs to quantum programs}. QuCheck provides a significant level of customisability, allowing for various inputs such as the insertion of random sub-circuits as inputs into a larger quantum program. 

\section{Property-Based Testing Quantum Programs}\label{property-based-testing-for-quantum-programs-secion} 
Property-based testing consists of the specification properties of the program under test, comprised of inputs, preconditions, operations, and postconditions. The property-based testing methodology generates a diverse array of inputs that satisfy the preconditions. These inputs are then passed to the program (circuit), which is executed, and the postconditions are checked against the results. Importantly, this approach does not require the user to manually identify specific inputs that could cause failures. Instead, properties serve as abstract specifications, allowing for the automated exploration of a wide range of input scenarios.

The application of property-based testing to quantum programs allows for:
\begin{itemize}
    \item \textbf{Handling diverse quantum inputs:} Quantum programs may receive oracle circuits or quantum states as input, comprised of multiple qubits in superposition or entanglement. property-based testing allows for the automatic exploration of these states and oracle circuits, without the necessity to manually specify them. 
    \item \textbf{Catching edge cases:} Because of this input diversity, it may be difficult for a developer to manually identify edge cases that cause the program to fail. property-based testing mitigates this problem through the generation of inputs that may be missed by a developer when designing tests.
    \item \textbf{Flexible Testing for Parameterized Circuits:} Quantum circuits are often parameterized, where their size or structure depends on input parameters, such as an integer specifying the number of qubits or gates. Property-based testing enables testing such parameterized circuits by allowing properties to be evaluated against a range of circuits generated from different parameter values. This removes the need to create and maintain tests for each potential circuit size or structure (i.e. different amounts of Grover diffusion operators applied in a Grover's algorithm circuit). 
    % \item \textbf{Regression Testing:} The property-based tests may be directly applied as regression tests.
    \item \textbf{Facilitation of Test-Driven Development:} In test driven development (TDD), developers first implement tests for their programs, then work on the implementation until they fail. However, it is challenging to construct test suites that catch all edge cases for quantum programs. Using an incomplete or incorrect suite of tests during TDD, leads to the development of an incomplete, or faulty program. Property-based testing avoids this scenario, as developers consider the broader semantics of the programs, and encode them into properties, rather than identify specific test cases.
\end{itemize}

\section{QuCheck}\label{QuCheck_section}

\subsection{Architecture}

\begin{figure}
    \vspace{10px}
    \centering
    \includegraphics[width=1\linewidth]{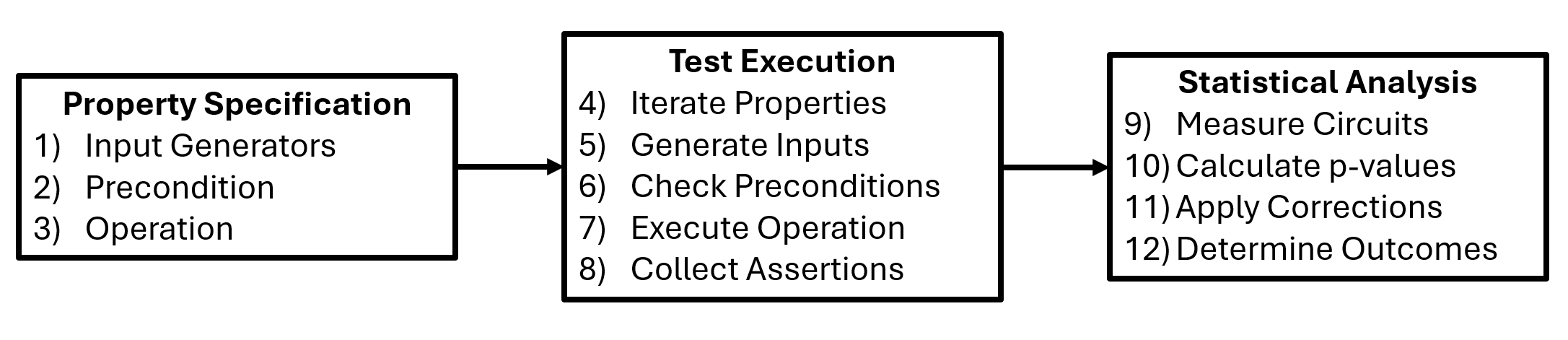}
    \caption{Major components of the framework.}
    \Description[The figure displays the three main components in separate boxes, with arrows pointing from "Property Specification", to "Test Execution" to "Statistical Analysis".]{The figure displays the three main components in separate boxes, with arrows pointing from "Property Specification" (with subheadings: "input generators", "precondition", "operation"), to "Test Execution" (with subheadings: "Iterate properties", "generate inputs", "check preconditions", "execute operation", "collect assertions") to "Statistical Analysis" (with subheadings: "measure circuits", "calculate p-values", "apply corrections", "determine outcomes").}
    \label{fig:architecture}
\end{figure}

The QuCheck architecture is based on three key components: property specification, test execution, and statistical analysis, as shown figure \ref{fig:architecture}.

\begin{itemize}

\item \textbf{Property Specification:} Encompasses the tools provided by QuCheck that aid the specification of properties, such as input generation, precondition specification, and assertion specification. 

\item \textbf{Test Execution:} Generates inputs using the provided input generators, evaluates them against preconditions, and executes the operations function defined in the property, which then collects and forwards assertions to the statistical analysis component.  

\item \textbf{Statistical Analysis:} Handles circuit optimization and execution, verifies both statistical and non-statistical assertions, and applies statistical corrections.  

\end{itemize}

\subsection{Property Specification}

Our framework streamlines the process of specifying and implementing property-based tests by providing tools and helper functions to aid in converting conceptual properties into executable code.

In QuCheck, properties are specified using four core elements, which inherit from a base property class provided by the framework. These elements include:

\begin{itemize}
    \item \textbf{Input generation} through the selection of a set of deterministic input generators that are needed for the property 
    \item \textbf{Precondition verification} by checking the inputs that have been generated through the generators
    \item \textbf{Operation execution} using the verified inputs for initialisation or inserting oracles.
    \item \textbf{Postcondition specification} by inserting assertions in the operation body
\end{itemize}

\subsubsection{Property Example}

An example of a property that has been specified using QuCheck can be seen in Listing \ref{lst:Property_example}. This property tests quantum teleportation by verifying that the state of the qubit in the target register after teleportation matches a qubit initialised to the same state. The \texttt{get\_input\_generators} function receives a list of input generators. In this case, \texttt{RandomState} is a wrapper around Qiskit's \texttt{random\_statevector} function, which samples from the uniform distribution induced by the Haar measure, meaning that every pure state is equally likely to be drawn. The parameter passed to the \texttt{RandomState} input generator indicates the size of the state to generate, in qubits.

No preconditions are required for this property, therefore, this \texttt{preconditions} function always returns \texttt{True}. An example usage of the \texttt{precondition} function is in the properties used to test Grover's algorithm, where the randomly generated Grover's oracle is checked to ensure that more (or less) than half of all possible states are marked. 

In the \texttt{operations} function, two circuits are initialised with the randomly generated state on register zero: \texttt{qc} and \texttt{qc2}. The quantum teleportation circuit is appended into \texttt{qc}, while \texttt{qc2} remains in its initialised state. The final step is to use \texttt{assert\_equal} to compare the second register of \texttt{qc} (which underwent teleportation) with register zero of \texttt{qc2} (the unaltered initialization).

\vspace{1em}
\begin{minipage}{0.9\linewidth}
\begin{small}
\begin{lstlisting}[
  caption=Code example for a quantum teleportation property in QuCheck,
  label={lst:Property_example},
  basicstyle=\scriptsize\ttfamily,  % smaller font
  aboveskip=2pt,                      % reduce space above the listing
  belowskip=2pt                       % reduce space below the listing
]
class TeleportationOutputEqualToInput(Property):
    # specify the inputs that are to be generated
    def get_input_generators(self):
        state = RandomState(1)
        return [state]

    # specify the preconditions for the test
    def preconditions(self, q0):
        return True

    # specify the operations to be performed on the input
    def operations(self, q0):
        qc = QuantumCircuit(3)
        qc.initialize(q0, [0])
        qt = quantum_teleportation()
        
        # stitch qc and quantum_teleportation together
        qc = qc.compose(qt)

        # initialise qubit to compare to:
        qc2 = QuantumCircuit(1)
        qc2.initialize(q0, [0])
        self.statistical_analysis.assert_equal(qc, [2], qc2, [0])
\end{lstlisting}
\end{small}
\end{minipage}

\subsubsection{Input Generators}

Properties require the specification of a range of inputs, which must be passed to the precondition check and body of the operation. To streamline this, we provide a predefined set of configurable, deterministic input generators, relevant for the formulation of common types of inputs to properties. Additionally, an interface is provided to allow user-defined input generators, which can also be created using a composite of predefined generators.

These input generators must be deterministic to ensure consistent input generation and enable developers to reproduce failing test cases during debugging. In QuCheck, this is achieved by requiring the input generators to use a seed, and ensuring that the same input is produced for each identical seed.

\subsubsection{Precondition}

The generated inputs are then passed to the precondition check, which verifies whether the randomly generated inputs satisfy the preconditions. QuCheck will attempt to generate an input that satisfies the precondition a configurable number of times, after which the property will time out and fail. 

\subsubsection{Operation}

The purpose of the operation function is to initialise the relevant circuits to evaluate, and specify the assertions that need to be verified. 

First, the operations function within a property receives the verified inputs, and uses them to set up the circuit or sub-circuit to be executed. For metamorphic properties, this function may construct two or more quantum circuits to compare their outcomes, avoiding the need for an oracle. The assertions to be verified are specified within this function and are passed to the statistical analysis component during execution. The post-conditions of the properties are verified using statistical assertions. These assertions are defined within the body of the operation function and stored in the statistical analysis component, where optimisations can be applied. 

\subsection{Test Execution}

The test execution component passes the verified inputs into the operation function of the property, which typically instantiates and initialises the circuit under test and specifies the assertions needed to verify the postcondition. After this process has completed for all properties, the assertions are collected and passed to the statistical analysis component for optimisation and execution. 

\subsection{Statistical analysis} 

The statistical analysis component iterates through all the collected assertions specified within the operations of properties, executing the referenced circuits and storing their results. A set of p-values for each generated input and property pair is calculated. Finally, a statistical correction is applied to adjust the significance level for each assertion to account for the multiple testing problem.

Before execution, the component applies an optimisation to minimise the number of circuits required, as discussed in Section \ref{optimisation_discussion}.

\subsubsection{Assertions}

Statistical assertions have been previously proposed \cite{Martonosi2019statisticalassertions} for the verification and validation of quantum states, these have been thus applied in \cite{DBLP:conf/icse/Honarvar0N20}. QuCheck contains a suite of predefined assertions that can be applied within the operation function that must be specified by each property. The common feature shared between all of the predefined assertions is the need to specify at least one quantum circuit, and a list of its qubit indexes that need to be measured.

Since performing full tomography on a set of qubits is computationally intensive, the predefined assertions within QuCheck perform measurements on the X, Y, Z basis for each qubit in the quantum state, comparing the marginal distribution of outcomes of one qubit at a time. This removes the assertions' ability to distinguish between different entangled states, yet drastically lowering the computational overhead gathering measurement results for the assertions. As we will demonstrate in this paper, this is an effective method for detecting faults. 

\begin{itemize}

\item \textbf{AssertEqual:} Receives two quantum circuits and two sets of qubit indices to compare with each other, as well as optionally the basis to compare, which defaults to X, Y, Z, but can be limited if necessary. 

Fisher's exact test is applied to the contigency tables constructed outcome distribution of each qubit and basis to determine the equality of the qubit's state. If any p-value returned from the test is below the adjusted significance level (from the statistical correction), the null hypothesis that the states are equal is rejected.  

\item \textbf{AssertDifferent:} The same as AssertEqual, except when verifying the p-values from the statistical tests, rather than rejecting the null hypothesis if any p-value is below the significance level, it is rejected if all p-values are above the significance level.

\item \textbf{AssertEntangled:} This assertion checks for entanglement within a circuit in a given basis between a provided set of qubits, assuming that the subset is maximally entangled. This is implemented by verifying whether the outcomes of the measured qubits consist of two complementary outcomes, such as $| 000 \rangle$ and $| 111 \rangle$, or $| 101 \rangle$ and $| 010 \rangle$. This approach scales linearly with number of qubits, but is limited to detecting entanglement within a single basis. It assumes that the subset of qubits passed to the assertion is maximally entangled in that basis. 

\item \textbf{AssertSeparable:} Checks for entanglement between the specified set of qubits and basis, then negates the result.

\item \textbf{AssertProbability:} For a given basis, asserts that the probability of measuring $| 0 \rangle$ for a set of qubits matches with a set of probabilities. 

\item \textbf{AssertMostFrequent:} No statistical test is applied for this assertion, we check that the most frequent outcome when measuring a provided quantum circuit is equal to the outcome provided.

\end{itemize}

Listing \ref{lst:Property_example} provides an example of using AssertEqual to verify that the states of a given set of qubits are equal in two quantum circuits.

\subsubsection{Measurement optimisation}\label{optimisation_discussion}

Multiple properties are evaluated in a property-based test suite, and some may require the evaluation of identical quantum circuits. In such cases, duplicate measurements (i.e., same basis and qubit registers on the same circuit) need to be performed only once, we describe the process below:

Before performing any measurements, the statistical analysis component collects all assertions by executing each properties' operation method, from which, the quantum circuits, measurement registers, and basis to measure are extracted and stored. The component then identifies and removes duplicate circuits and determines what measurements need to be performed. For each unique circuit, the statistical analysis component starts by making a copy, it then iterates through the list of measurements (and basis changes) to attach, adding each one to the circuit if it does not conflict with an already inserted measurement. If a conflict occurs, the measurement is added to a new, or existing copy of the circuit where the conflict does not occur. Finally, a pass is done to verify that no duplicate circuits are present after the basis changes and measurements are inserted.

It should be noted that when the same set of input generators is passed to multiple properties, the test runner will attempt to reuse the same seeds, resulting in identical inputs across these properties. This may cause identical circuits to be executed (if preconditions fail, additional seeds are generated), which are then deduplicated and optimised as described above, allowing measurements to be reused for the verification of properties.

\subsubsection{Statistical correction}

property-based testing generates multiple test cases each of which call a statistical assertion to verify their post-condition. The probability of making a Type I error increases when applying multiple statistical tests consecutively. To address this, the family-wise error rate is corrected through the application of a statistical correction such as the Holm-Bonferroni correction.  

To apply this correction, the p-values from each statistical test need to be collected. To do this, a fixed number of inputs are generated, tested as a batch, and evaluated simultaneously. Unlike QSharpCheck, this approach does not test each input individually until a property fails (or a set number of inputs has been tested). This is discussed further in Section \ref{label:limitations}.

\subsection{Limitations}\label{label:limitations}
One limitation of this approach is the inability adaptively control the number of inputs based on the test results, a set number of inputs to generate must be first specified and tested, as the assertion check is only performed after all the test cases have been evaluated. Practically, this may lead to more circuits being executed and measured than necessary to verify a given property. This design choice was made to allow for the application of a statistical correction, which requires knowledge of all the statistical tests that are performed. 

Several issues arise from property-based testing's input generation approach compared to fixed test case methods. For some programs, specific, carefully chosen inputs are necessary to validate program behaviour under known conditions. Furthermore, exploring the input space through random generation is more resource-intensive than using optimised, pre-planned input sets that maximise coverage and adequacy metrics with minimal inputs. Constructing input generators for complex inputs, such as the oracles or sub-circuits within quantum circuits, requires significant domain knowledge and effort.   

As shown in our experiments, defining an adequate suite of properties for complex systems essential for property-based testing to be effective. However, this is not a trivial task, as it requires thorough knowledge of the program under test to design multiple properties whose combined coverage adequately represents the program's semantics.

Verifying postconditions for quantum programs remains challenging due to the trade-off between accuracy and performance. Various approaches can be used implement assertions for this purpose, ranging from full tomography, which is accurate but resource intensive for large quantum systems, to evaluating the individual qubits outcomes on a limited number of bases. Other techniques, such as classical shadows \cite{Preskill}, can efficiently assert a limited subset of states, though they come with their own limitations. 

\section{Experiment Design}\label{experiment-design}

Two sets of experiments were performed to address our research questions. The first set evaluated QuCheck in a isolation to evaluate the \textit{effectiveness} of property-based testing under varying configurations of \textit{thoroughness}. The second set involved a quantitative comparison between QuCheck and QSharpCheck, where two case studies from the original experiment were repeated, with newly translated mutants. Below, we provide details on the mutation analysis used in our experiments, which was used to measure the effectiveness of the testing techniques. Furthermore, we describe the independent variables used to define the thoroughness of the test suite, and the quantum programs used as our subject systems.

\subsection{Mutation Analysis}

Mutation analysis is employed to assess the effectiveness of a testing suite. The general technique is to apply small changes to a correct implementation of a program to create a set of "mutants". The test suite to be evaluated is then applied to each mutant, and the test outcome is recorded. The effectiveness of the test suite is subsequently determined by calculating the mutation score, which is the percentage of mutations that failed the tests over the total number of mutations. However, achieving 100\% mutation score is unlikely with a large set of mutants, as some mutant versions of the program may be semantically equivalent to the original program (known as equivalent mutants), making them to be impossible to be detected by a test suite. 

QMutPy was used in the experiments to generate ten mutants for each of the five quantum programs in our case studies. However, some mutants generated by the tool were syntactically incorrect. In such cases, new mutants were generated until no more syntactic errors occurred. 

To measure the \textit{false positive rate} of the property-based tests, a separate set of equivalent mutants was intentionally generated. This was done by randomly selecting a set of gate identities, inserting them into the case study implementation of the circuit. Once the mutants were generated, QuCheck was applied to them, the outcomes were recorded, and are listed in Section \ref{results}. 

\subsection{Subject systems}

For the evaluation of QuCheck, ten mutations of each of the five diverse quantum programs were generated. These subject systems were chosen for their variation in gate depth, circuit width, and usage of oracles. Specifically, Deutsch Jozsa and Grover's algorithms were included to evaluate how QuCheck handles programs that receive oracles as input. Quantum phase estimation was selected for its more complex inputs involving a unitary operator and a state vector. Additionally, quantum fourier transform and teleportation were also included as simpler algorithms to test. Finally, each algorithm except quantum teleportation is configurable, allowing their width and depth vary depending on their inputs. This enables us to evaluate how well property-based testing performs for algorithms that scale in size. 

\begin{itemize}
  \item \textbf{Quantum Fourier Transform (QFT):} The Quantum Fourier transform algorithm efficiently applies the Fourier transform to the amplitudes of any inputted state. The algorithm is applied within many other programs, such as Shor's algorithm and quantum phase estimation.  
  \item \textbf{Quantum Phase Estimation (QPE):} QPE estimates the value of $\varphi$ when given a unitary operator $U$, and an eigenvector of the unitary $|u\rangle$ that has an eigenvalue of $e^{2 \pi i \varphi }$. A well known application is its use within Shor's algorithm to find the order $r$ of $a^{r} \equiv 1$ mod $N$, where $a$ is co-prime with $N$, which can be then be used to efficiently find the factors of a large number $N$.  
  \item \textbf{Quantum Teleportation (QT):} Quantum teleportation transfers the quantum state of a single qubit $| \psi \rangle $ to another qubit. This process requires the transmission of one qubit from an entangled pair, as well as a means to transfer two bits of classical information, and has been experimentally realised with an average fidelity of 0.80 ± 0.01 over a distance of up to 1400 kilometers \cite{Ren2017-yc}. Quantum teleportation and its derivatives have applications in quantum networks \cite{Hermans2022}, allowing for the reliable transfer of quantum information over long distances.
  \item \textbf{Grover's algorithm:} Grover's algorithm, often called the quantum search algorithm, can be applied to find the inputs, when given the solution to a function (find $x$ where $f(x) = 1$). The algorithm repeatedly increases the amplitudes of the indexes (input values) that satisfy the given black box function by repeating the application of the Grover diffusion operator before performing measurements. This algorithm can provide a quadratic speedup to NP-complete problems \cite{Nielsen2012, hamiltonianCycleSpeedup, npCompleteSpeedup}.   
  \item \textbf{Deutsch Jozsa's algorithm:} The Deutsch Jozsa algorithm is able to determine whether a black box oracle applies either a constant or balanced function to the lower register using only a single query. 
\end{itemize}

\subsection{Procedure}

For the rest of this section, we define the experimental procedure and rationale to answer each research question:

\begin{itemize}
    \item \textbf{RQ1:} How does the \textit{thoroughness} of a property-based testing suite affect the ability to identify faults in the quantum programs?
    \item \textbf{RQ2:} Is property-based testing \textit{effective} for the identification of faults in quantum programs?
    \item \textbf{RQ3:} How does QuCheck compare to QSharpCheck?
\end{itemize}

\subsection{Experimental setup for RQ1 and RQ2}

To answer \textbf{RQ1} and \textbf{RQ2}, we define the metrics used to evaluate the effectiveness of the property-based testing suite in detecting faults, as well as methods to adjust its thoroughness for comparison.

\begin{itemize}
    \item \textbf{Effectiveness:} The mutation score was the primary metric used to evaluate effectiveness, though it is assessed differently depending the set of mutants being tested. After the initial mutant generation with QMutPy, manual inspection was performed to create two sets of mutants: one containing only \textit{equivalent mutants} and the other containing only \textit{QMutPy} mutants. For the former, the mutation score indicates the rate of false positives, which need to be minimised; conversely, for the semantically different set of mutants, the mutation score indicates the ability to detect potential faults. The execution time of the property-based tests for each configuration was also employed to evaluate the cost-benefit trade-off in increasing the test suite thoroughness.   
    \item \textbf{Thoroughness:} The thoroughness of the test suite was varied by adjusting three variables: the number of properties considered in the property-based tests, the number of inputs generated for each property, and the number of measurement shots. All combinations of these independent variables (Table. \ref{independent-variables}) were used to evaluate the technique.    
\end{itemize}

\begin{table}[h]
    \vspace{6px}
    \begin{tabular}{l}
    \begin{tabular}{p{3cm}|>{\centering\arraybackslash}p{0.6cm}>{\centering\arraybackslash}p{0.6cm}>{\centering\arraybackslash}p{0.6cm}} 
         \hline
         Number of Properties &  1&  2& 3\\
         \hline
    \end{tabular}

    \vspace{6px}
    \\

    \begin{tabular}{p{3cm}|>{\centering\arraybackslash}p{0.6cm}>{\centering\arraybackslash}p{0.6cm}>{\centering\arraybackslash}p{0.6cm}>{\centering\arraybackslash}p{0.6cm}>{\centering\arraybackslash}p{0.6cm}>{\centering\arraybackslash}p{0.6cm}>{\centering\arraybackslash}p{0.6cm}}
        \hline
        Number of Inputs & 1 & 2 & 4 & 8 & 16 & 32 & 64\\
        \hline
    \end{tabular}

    \vspace{6px}
    \\

    \begin{tabular}{p{3cm}|>{\centering\arraybackslash}p{0.6cm}>{\centering\arraybackslash}p{0.6cm}>{\centering\arraybackslash}p{0.6cm}>{\centering\arraybackslash}p{0.6cm}>{\centering\arraybackslash}p{0.6cm}>{\centering\arraybackslash}p{0.6cm}>{\centering\arraybackslash}p{0.6cm}>{\centering\arraybackslash}p{0.6cm}>{\centering\arraybackslash}p{0.6cm}}
         \hline
         Number of Shots & 12 & 25 & 50 & 100 & 200 & 400 & 800 & 1600 & 3200\\
         \hline
    \end{tabular}
    \end{tabular}
    \vspace{6px}
    \caption{Independent variables considered.}\label{independent-variables}
\end{table}

\begin{figure}[h]
    \centering
    \includegraphics[width=0.95\linewidth]{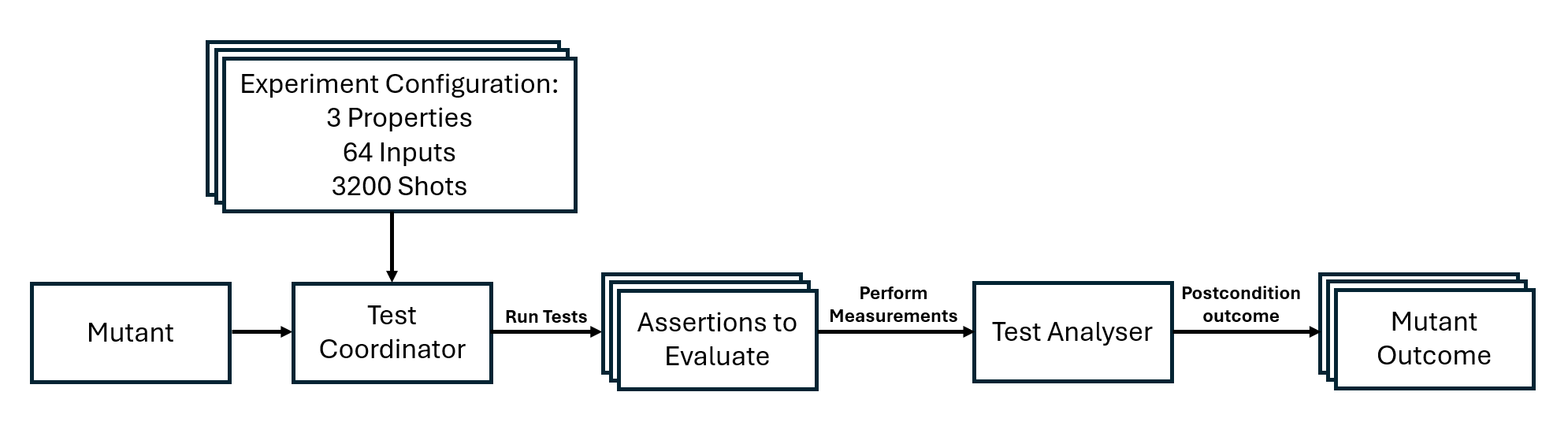}
    \caption{Experimental setup.}
    \Description[Diagram portraying the flow from mutant to its outcome depending on experimental configuration.]{Diagram portraying the flow from mutant to its outcome depending on experimental configuration, the experimental configurations and the mutant are passed to the test coordinator, which calculates which assertions need to be executed, passes them to the test analyser which determines their outcome.}
    \label{fig:exp-setup}
\end{figure}

QuCheck was configured to execute each of these configurations against every mutant of the program being tested. A mutant was deemed "killed" if any of the property-based tests failed during execution. Figure \ref{fig:exp-setup} illustrates this experimental process, demonstrating the flow from configuration generation through mutant testing to result analysis.

For the evaluation of the relationship between the thoroughness and effectiveness of the test suite (\textbf{RQ1.1} to \textbf{RQ1.3}), we systematically tested all possible combinations of these variables from Table \ref{independent-variables}, henceforth referred to as configurations.

To determine the overall effectiveness of property-based testing for quantum programs (\textbf{RQ2}), we considered the mutation score and execution time of only the configurations with the \textbf{largest two values for each independent variable} (2, 3 properties, 32, 64 inputs, and 1600, 3200 measurement shots). 

\subsection{Experimental setup for RQ3}

To the best of our knowledge, the only other published property-based test framework for quantum programs is QSharpCheck \cite{DBLP:conf/icse/Honarvar0N20}. The process for determining the qualitative differences (\textbf{RQ3.1}) between QSharpCheck and QuCheck is described in section \ref{qualitative-differences}, the following defines the experimental setup to evaluate the differences in effectiveness and execution time between the frameworks (\textbf{RQ3.2}). To perform this evaluation, we utilise the largest two configuration of independent variables as in \textbf{RQ2}, except testing only one property. Since QuCheck measures three bases rather than one in its implementation of the properties, we adjusted the number of measurements for QSharpCheck to achieve parity in the \textbf{total number of measurements}. Specifically, QSharpCheck was tested with three times the number of measurements per basis (i.e., 9600 instead of 3200) to match the total number of measurements performed by QuCheck. For comparison, we also included a configuration which matches the number of measurements per basis, but results in fewer total measurements for QSharpCheck.

\subsubsection{Conversion of mutants:} To ensure the fairness of the comparison, the same mutants that were generated for the evaluation of QuCheck (\textbf{RQ1-2}) were also used to compare QSharpCheck. Since  QSharpCheck's framework's programming language is different, we needed to translate the Qiskit mutants into Q\#. This was achieved by first converting each Qiskit mutant into QASM 2.0 directly through Qiskit's API, then using Quantastica's QConvert to translate from QASM 2.0 into Q\# 0.10. This automatic translation often produced syntactically incorrect code, which was manually refactored for its execution in QSharpCheck. 

\subsubsection{Conversion of properties and assertions:} Unfortunately, due to the lack of features, it was not possible to convert the majority of the QuCheck properties into QSharpCheck properties. For example, in QuCheck, AssertEqual is available for multiple qubits, while QSharpCheck limits AssertEqual to single qubits. We also encountered failures when attempting to use a portion of the provided assertions, preventing us from replicating many properties. We expect changes to the Q\# compiler may have contributed to these.

\subsubsection{Conversion of configurations:} QSharpCheck's statistical analysis configuration requires three variables: inputs, experiments, and shots. The total measurements per input is equal to the number of experiments multiplied by shots. For example, to perform a a total number of 3200 measurements, we set experiments to 40 and shots to 80.

\subsection{System configuration}

The experiments were performed in Qiskit 1.0.1 using a windows 11 machine (R7 5700X, RTX 3060ti, 16GB RAM). 

\section{Results}\label{results}

\subsection{RQ1: Impact of test suite thoroughness on effectiveness}\label{RQ1}

\begin{figure}[h]
    \centering
    \includegraphics[width=0.95\linewidth]{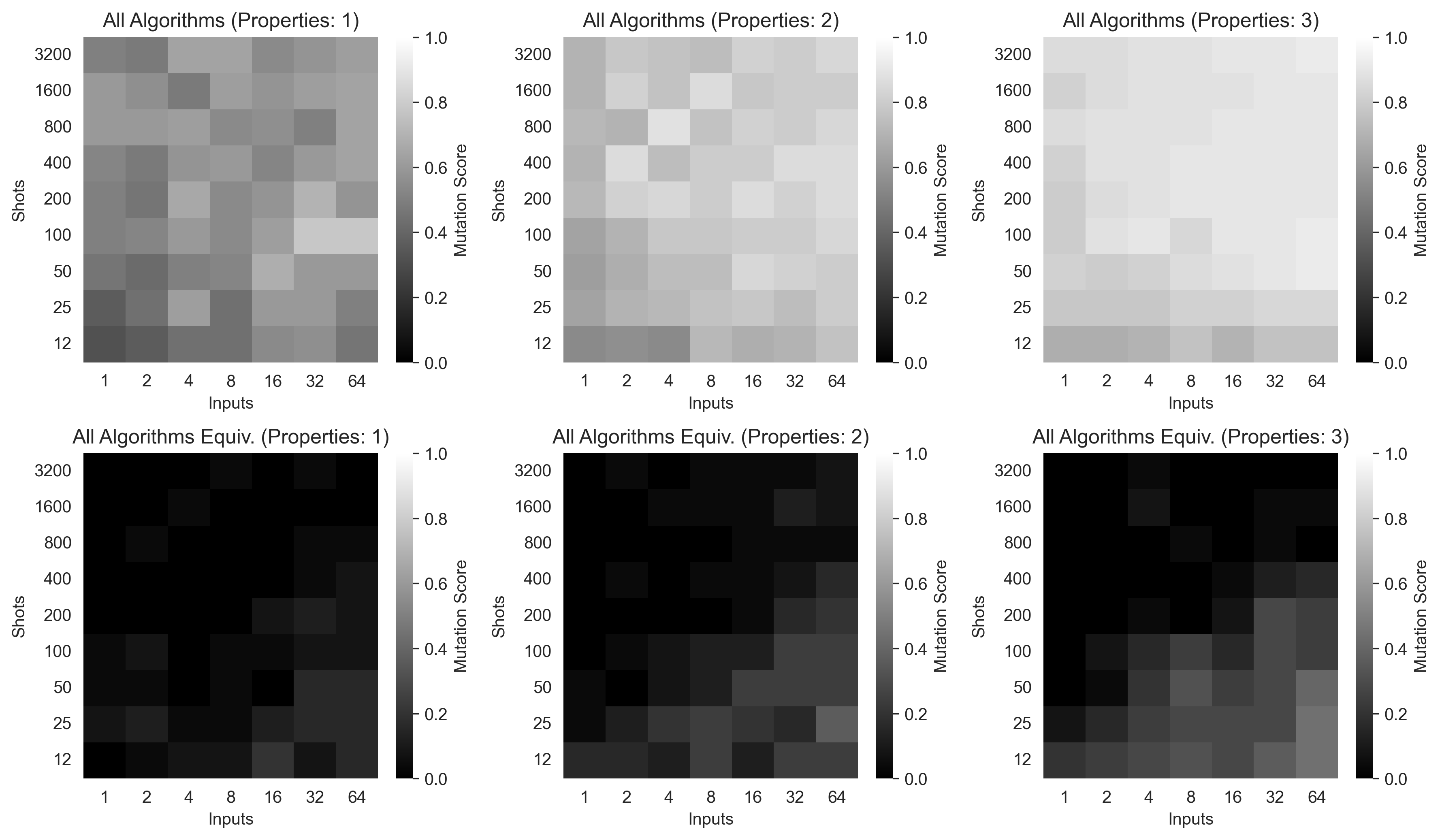}
    \caption{Heatmaps illustrating the impact of property count, input size, and measurement quantity on mutation scores. The top row represents the average mutation scores for QMutPy-generated mutants across all algorithms, and the bottom row, the average mutation scores for equivalent mutants.}
    \Description[A row of 3 heatmaps portraying the mutation score of different test configurations, portraying higher mutation score with higher number of inputs and measurements, overall higher score with more properties, more noise in the heatmaps with less properties. Below is another row of 3 heatmaps portraying the mutation score for different equivalent mutant configurations, it appears as if more inputs and properties causes a higher mutation score, but if a higher number of measurements is a applied, this effect is lost.]{A row of 3 heatmaps portraying the mutation score of different test configurations, portraying higher mutation score with higher number of inputs and measurements, overall higher score with more properties, more noise in the heatmaps with less properties. Below is another row of 3 heatmaps portraying the mutation score for different equivalent mutant configurations, it appears as if more inputs and properties causes a higher mutation score, but if a higher number of measurements is a applied, this effect is lost.}
    \label{fig:average_heatmap}
\end{figure}

Figure \ref{fig:average_heatmap} shows six heatmaps that illustrate the effects of increasing property count, input size, and number of measurement shots on the average mutation scores across all algorithms. The top row corresponds to QMutPy-generated mutants, while the bottom row represents equivalent mutants. The columns, from left to right, depict an increasing numbers of properties, from one to three.

Two key patterns emerge from these heatmaps. First, for \textbf{QMutPy mutants} (top row), there is a clear trend of higher mutation scores in the top right region of each heatmap, indicating that a larger number of inputs and property counts lead to an increase in the mutation score. This trend is more pronounced when three properties are evaluated. This increase in mutation score for the QMutPy mutants is supported by a statistically significant positive correlation observed in the individual analysis of each independent variable, as investigated in RQ1.1, RQ1.2 and RQ1.3.

In contrast, for \textbf{equivalent mutants} (bottom row), a different overall pattern emerged. Lower mutation scores are concentrated at higher number of measurement shots rather than at lower number of measurements and inputs. Although a positive correlation with the number of inputs and properties remains, \textit{a stronger negative correlation} with the number of measurement shots was observed. Furthermore, when restricting the number of properties in Figure \ref{fig:num_properties_line_high_measurements} to utilise the highest three configurations of measurements, \textit{this positive correlation disappears}. 

\begin{tcolorbox}[colback=blue!5!white,colframe=blue!75!black, enlarge top by=0.2cm, enlarge bottom by=0.2cm, title=\textbf{Answer to RQ1:} How does the \textit{thoroughness} of a property-based testing suite affect the ability to identify faults in the quantum programs?]

Increasing the thoroughness of the property-based testing suite \textbf{increases the mutation score for the QMutPy mutants}, enhancing the ability to identify faults in quantum programs. At the same time, increasing the number of measurements performed per input \textbf{decreases the mutation score for the set of equivalent mutants}, thus reducing the likelihood of false positive results. 

\end{tcolorbox}

In the subsections below, we answer the sub-questions (RQ1.1-RQ1.3), illustrating the impact of each independent variable on the mutation score of each algorithm. In Figures \ref{fig:num_properties_line}-\ref{fig:num_shots_line}, the dotted lines represent the mean values for each individual algorithm, stratified by the independent variable plotted on the x-axis. The solid red line represents the mean value over all algorithms and measurement configurations, with the vertical bars representing the 95\% confidence interval of the true population mean. The Spearman's rank correlation coefficient was used to assess the relationship between the variables, which is suited to the monotonic (and sometimes non-linear) trend observed. 

\subsubsection{RQ1.1: Effect of Property Count per Program}

\begin{figure}[H]
    \centering
    \includegraphics[width=1\linewidth]{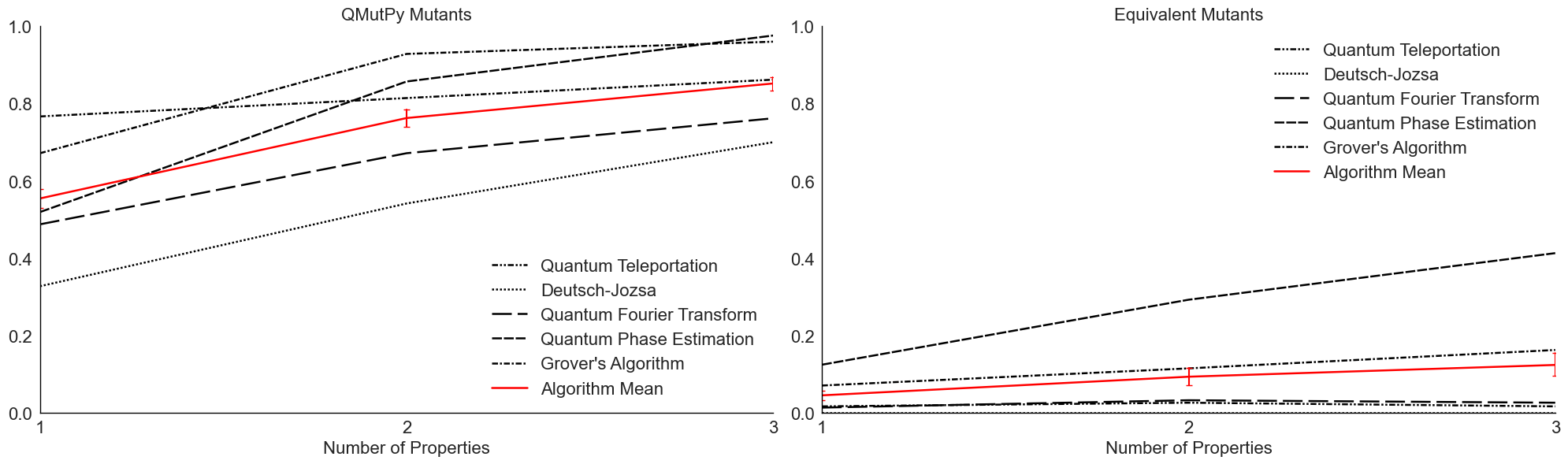}
    \caption{Effect of \textit{number of properties} on mutation score.}
    \Description[Two graphs portraying an increasing trend with number of properties and mutation score.]{Two graphs portraying an increasing trend with number of properties and mutation score.}
    \label{fig:num_properties_line}
\end{figure}

The mutation score of mutants generated through \textbf{QMutPy} showed a strong positive correlation ($r$ = 0.532, $p$ = 2.75e-70), indicating a statistically significant relationship between the number of properties evaluated in the test suite and the mutation score. This trend can be seen in Figure \ref{fig:num_properties_line}. The median mutation scores for tests evaluating 1, 2 and 3 properties were 0.6, 0.8, and 0.9, respectively, with corresponding standard deviations of 0.223, 0.202, and 0.157. The reduction in standard deviation highlights the increased consistency of the mutation score as the number of properties evaluated in the test suite (thus program semantics covered by tests) also increases. 

\begin{figure}[H]
    \centering
    \includegraphics[width=1\linewidth]{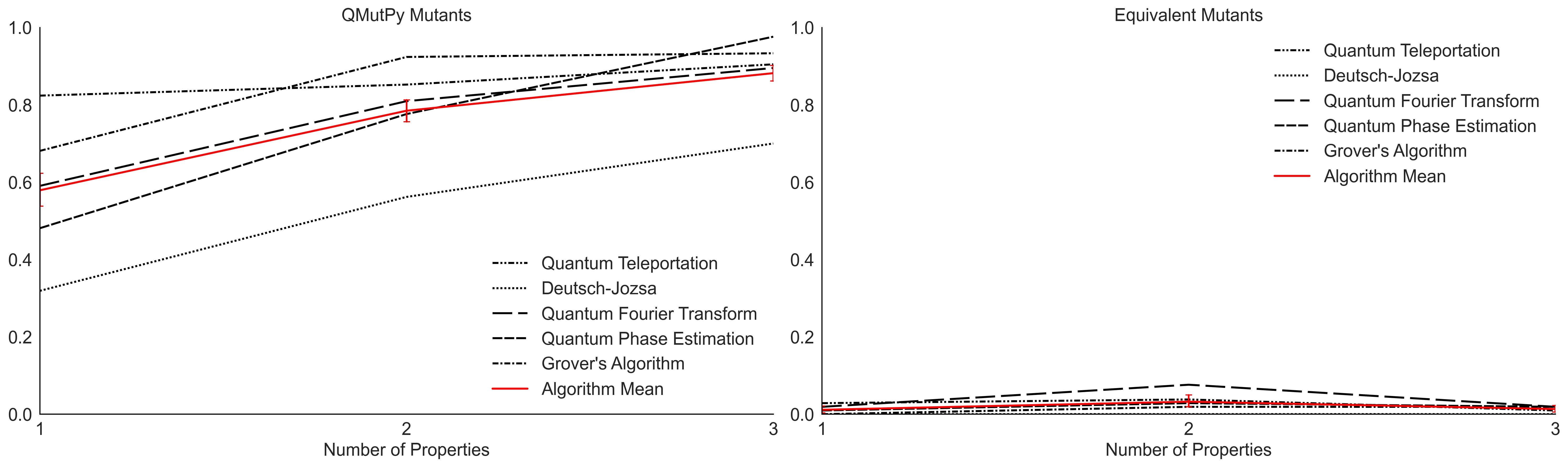}
    \caption{Effect of \textit{number of properties} on mutation score with 800, 1600, and 3200 measurement shots.}
    \Description[Two graphs, one portraying an increasing trend with number of properties and mutation score, the equivalent mutant graph showing no corellation between number of properties and mutation score]{Two graphs, one portraying an increasing trend with number of properties and mutation score, the equivalent mutant graph showing no corellation between number of properties and mutation score}
    \label{fig:num_properties_line_high_measurements}
\end{figure}

A weak positive correlation ($r$ = 0.077, $p$ = 0.019) was observed between the number of properties and the mutation score for \textbf{equivalent mutants}, indicating a slight increase in false positive results when more properties were evaluated. Among the five algorithms studied, \textit{quantum phase estimation} showed the most pronounced increase in false positives as properties increased. The overall trend of rising false positives is primarily caused by the insufficient measurement shots in test configurations. This was verified by averaging only test configurations with measurement shots set to 800, 1600, and 3200 (Fig. \ref{fig:num_properties_line_high_measurements}); no statistically significant correlation was then be found with the number of properties ($r$ = 0.014, $p$ = 0.810). This suggests a dependency between algorithmic constructs, the specific properties being tested, and the number of measurement shots required to minimise false positive results.

\begin{tcolorbox}[colback=blue!5!white,colframe=blue!75!black, enlarge top by=0.2cm, enlarge bottom by=0.2cm, title=\textbf{Answer to RQ1.1:} How does the \textit{number of properties per program} affect the ability to identify faults in the quantum programs?]

Increasing the number of properties in quantum property-based test suites significantly improves fault detection in quantum programs. While this initially raises the false positive rate, using an adequate number of measurement shots (which is algorithm-property dependent) mitigates this effect. 

\end{tcolorbox}

\subsubsection{RQ1.2: Effect of Number of Inputs per Property}

\begin{figure}[H]
    \centering
    \includegraphics[width=1\linewidth]{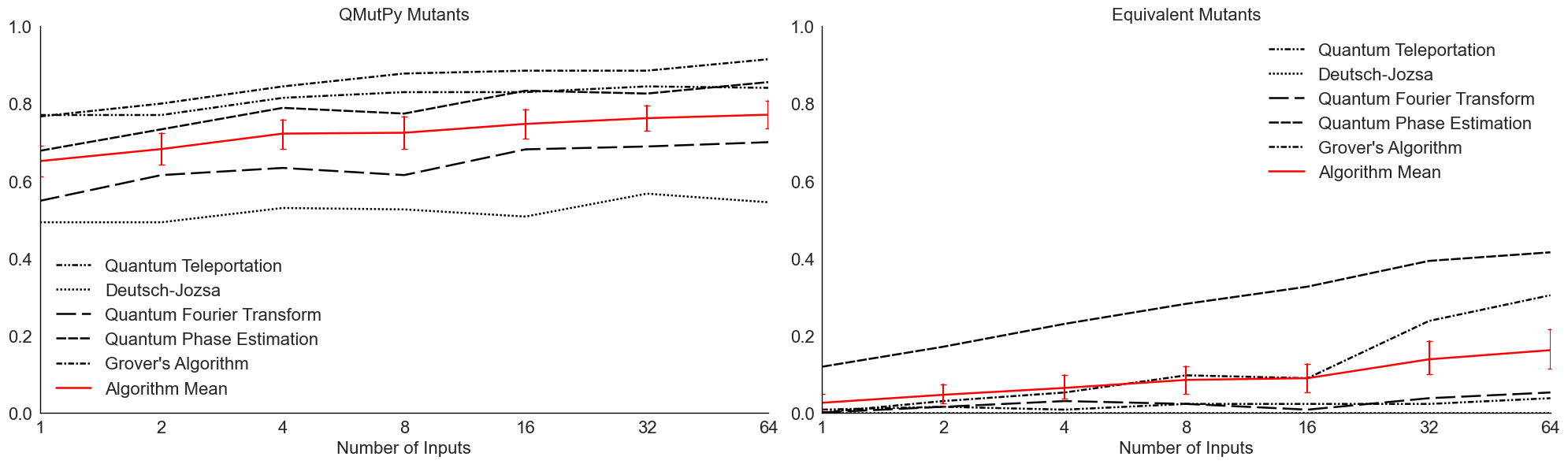}
    \caption{Effect of \textit{number of inputs} per property on mutation score.}
    \Description[Two graphs portraying an increasing trend with number of inputs and mutation score.]{Two graphs portraying an increasing trend with number of inputs and mutation score.}
    \label{fig:num_inputs_line}
\end{figure}

Analysis of Figure \ref{fig:num_inputs_line} revealed a weak positive correlation between the number of inputs generated within a property-based test and the mutation score ($r$ = 0.173, $p$ = 9.300e-8). The median mutation score observed was 0.7 when generating one input, and 0.8 for all other input configurations. The standard deviation ranged between 0.210 to 0.253 across the different number of inputs, without a clear decreasing pattern as seen when varying the number of properties.

The \textbf{equivalent mutant} set showed a similar trend to the QMutPy set; a weak positive correlation ($r$ = 0.227, $p$ = 1.640e-12) was observed. The median mutation score remained at zero for all configurations of number of inputs. However, The standard deviation of the mutation score increased by approximately 0.03 when doubling the number of inputs, ranging from 0.123 at one input, to 0.292 at 64 inputs, indicating a decrease in consistency of the mutation scores. As in RQ1.1, the \textit{quantum phase estimation algorithm} (Fig. \ref{fig:num_inputs_line}) had the highest false positive rate across the range of inputs.    

When considering only the subset of results containing \textit{three properties and three largest numbers of measurement shots}, the positive correlation between the number of inputs and the mutation score for the \textbf{equivalent mutant set} lost its statistical significance ($r$ = 0.076, $p$ = 0.439). Conversely, the correlation for \textbf{QMutPy mutants}' increased to $r$ = 0.239, $p$ = 0.014. 

\begin{tcolorbox}[colback=blue!5!white,colframe=blue!75!black, enlarge top by=0.2cm, enlarge bottom by=0.2cm, title=\textbf{Answer to RQ1.2:} How does the \textit{number of inputs per property} affect the ability to identify faults in the quantum programs?]

Increasing the number of inputs generated per property has a moderate impact on fault identification for QMutPy mutants, with a weak positive correlation observed for the equivalent mutants. However, With the most thorough configurations, property-based testing effectively leverages the increased number of inputs to detect more faults and reduce false positives.

\end{tcolorbox}

\subsubsection{RQ1.3: Effect of Measurement Shots}

\begin{figure}[H]
    \centering
    \includegraphics[width=1\linewidth]{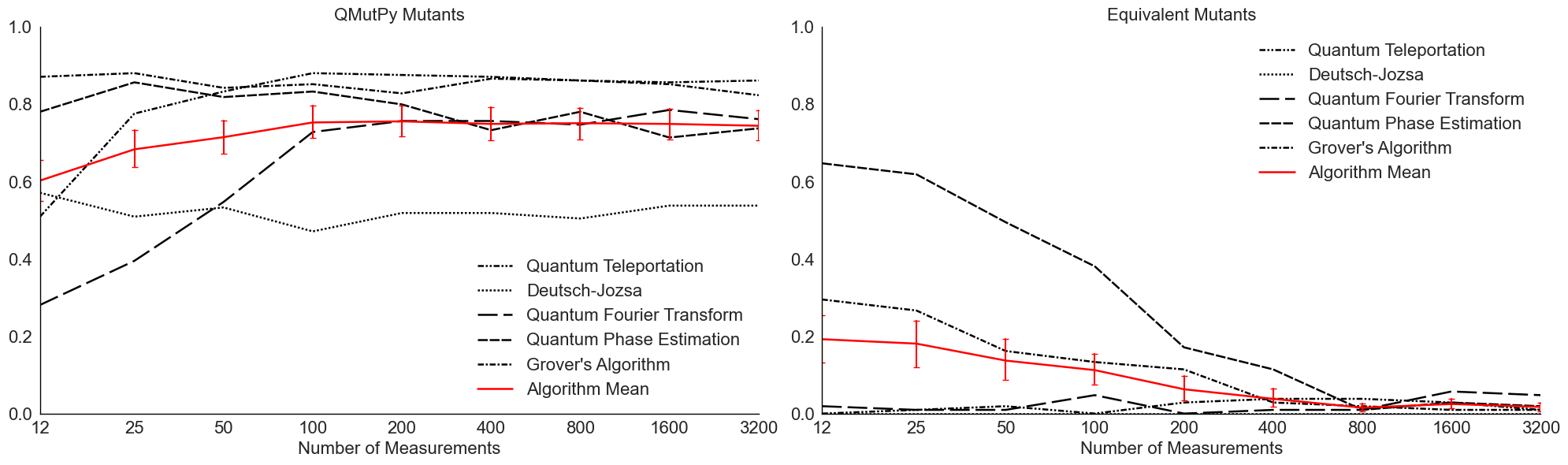}
    \caption{Effect of \textit{measurement shots} on mutation score.}
    \Description[Two graphs, the QMutPy mutant graph portrays an increasing trend until 200 measurements, which flattens. The equivalent mutant graph's mutation score decreases until 800 measurements, then flattens.]{Two graphs, the QMutPy mutant graph portrays an increasing trend until 200 measurements, which flattens. The equivalent mutant graph's mutation score decreases until 800 measurements, then flattens.}
    \label{fig:num_shots_line}
\end{figure}

A weak statistical correlation was observed between the number of measurements performed (Fig. \ref{fig:num_shots_line}) for each input in the property-based test ($r$ = 0.121, $p$ = 1.999e-4). The observed increase in mutation score occurred between 12 and 200 measurements, where the mean increased from 0.603 to 0.756, before plateauing. The median mutation score increased from 0.6 at 12 measurements to 0.7 at 25 and 50, and 0.8 for the remaining measurement values. The standard deviation of the mutation score decreased from 0.272 to 0.200 across the range of inputs.

The trend for \textbf{equivalent mutants} differed from the patterns observed with other variables. A statistically significant negative correlation ($r$ = -0.266, $p$ = 9.453e-17) was found between the number of measurements and mutation score. The median mutation score remained constant at zero across all measurements, while the consistency of the output improved, as shown by a reduction in the standard deviation of the mutation score from 0.327 to 0.063.

\begin{tcolorbox}[colback=blue!5!white,colframe=blue!75!black, enlarge top by=0.2cm, enlarge bottom by=0.2cm, title=\textbf{Answer to RQ1.3:} How does the \textit{number of measurement shots} affect the ability to identify faults in the quantum programs?]

Although increasing the number of measurement shots raises the mutation scores for QMutPy mutants, we observed a threshold at 200 measurements, beyond which there are diminishing returns. 

For equivalent mutants, the mutation score greatly decreases as the number of measurements increases, showing a significant negative correlation that can be leveraged to reduce the probability of false positives in property-based tests. Another threshold was observed at 800 measurements, where an increase to the number of measurements yields a negligible reduction in false positives.

\end{tcolorbox}

\subsection{RQ2: property-based testing effectiveness}

The aim with this research question is to evaluate whether property-based testing can effectively identify faults in quantum programs. To this end, only the two largest configurations of each independent variable were considered. As in section \ref{RQ1}, the primary measure of effectiveness being used is \textbf{mutation score}, which indicates the ability of the property-based test suite's ability to detect faults introduced by mutants. To assess the feasibility of running these tests, we also recorded the execution time of each algorithm in Table \ref{time-taken}.  

\begin{table}[H]
    \small
        \begin{tabular}{ c|c|c|c|c|c|c|c|c }
            \hline
            \multicolumn{3}{c|}{Configuration} & \multicolumn{5}{c|}{Algorithm Average Time Taken (s)} &  \\ 
            \hline
            Properties & Inputs & Measurements & QT & DJ & QFT & QPE & Grover & Mean \\
            \hline
            3  &  64  &  3200  &  6.71 &  3.66  &  45.10  &  49.05  &  120.00 & 44.90\\
            3  &  64  &  1600  &  4.93 &  3.14  &  31.57  &  43.18  &  115.09 & 39.58\\
            3  &  32  &  3200  &  3.45 &  1.86  &  16.06  &  26.29  &  58.19  & 21.17\\
            3  &  32  &  1600  &  2.54 &  1.59  &  15.64  &  22.50  &  53.49  & 19.15\\
            \hline
            2  &  64  &  3200  &  4.89 &  2.50  &  31.81  &  30.85  &  50.90  & 24.19\\
            2  &  64  &  1600  &  3.48 &  2.07  &  20.80  &  25.73  &  39.95  & 18.41\\
            2  &  32  &  3200  &  2.42 &  1.17  &  10.82  &  18.95  &  11.16  & 8.90\\
            2  &  32  &  1600  &  1.65 &  1.04  &  9.05   &  16.59  &  16.90  & 9.05\\
        \end{tabular}
        \caption{Execution time comparison of different configurations.}\label{time-taken}
\end{table}

\begin{figure}[H]
    \centering
    \includegraphics[width=0.95\linewidth]{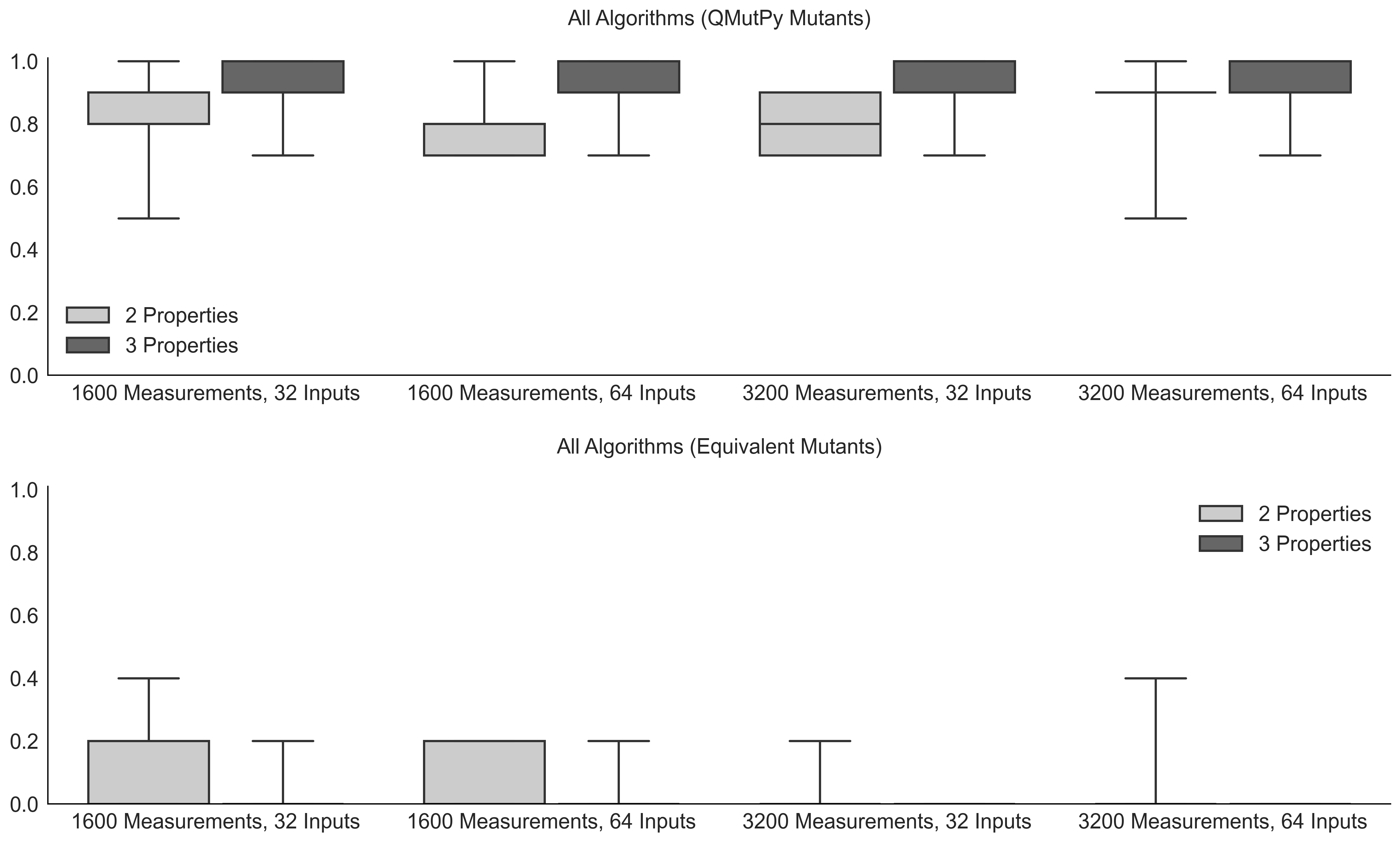}
    \caption{Effectiveness of property-based Testing on the most thorough configurations.}
    \Description[Two graphs, the QMutPy mutant graph portrays box plots for mutation score for the most thorough configurations, the box plots with more properties, inputs and measurements are placed higher in the axis. In the equivalent mutant graph, the box plots are placed lower in the axis, displaying a lower false positive rate.]{Two graphs, the QMutPy mutant graph portrays box plots for mutation score for the most thorough configurations, the box plots with more properties, inputs and measurements are placed higher in the axis. In the equivalent mutant graph, the box plots are placed lower in the axis, displaying a lower false positive rate.}
    \label{fig:boxplot_effectiveness}
\end{figure}

Figure \ref{fig:boxplot_effectiveness} presents two box plots that illustrate the relative mutation scores for the most thorough configurations. The upper box plot represents the mutation scores for \textbf{QMutPy-generated mutants}, while the lower one displays the scores for \textbf{equivalent mutants}. Configurations using three properties consistently outperformed those with two properties in terms of mutation score. For the QMutPy mutants, mean mutation scores ranged from 0.80 to 0.84 with two properties and increased to between 0.90 and 0.92 with three properties. In contrast, the mutation scores for equivalent mutants decreased, ranging from 0.04 to 0.12, and from 0 to 0.04 for two and three properties, respectively. 

The improved fault-detection capability came at the expense of execution time, though it remained feasible across all case studies. For the configuration with 3 properties, 64 inputs, and 3200 measurements, the execution times varied from 3.66 seconds for the Deutsch-Jozsa algorithm to 120.00 seconds for Grover's algorithm. This is in contrast to the average time taken for the weakest configuration, which ranged from 1.04 seconds to 16.90 seconds.

\begin{tcolorbox}[colback=blue!5!white,colframe=blue!75!black, enlarge top by=0.2cm, enlarge bottom by=0.2cm, title=\textbf{RQ2:} Is property-based testing \textit{effective} for the identification of faults in quantum programs?]

The high mutation scores observed across the most thorough configurations demonstrate a significant ability to detect faults in quantum programs, while maintaining feasible execution times across all tested properties and configurations. 

\end{tcolorbox}

\subsection{RQ3: Comparison of QuCheck and QSharpCheck}

In Section \ref{qualitative-differences}, we identify the qualitative differences between the two frameworks, finding QuCheck offers greater flexibility and extensibility in defining properties compared to QSharpCheck. Key differences include the programming language used, the format for property specification, the input generators and assertions, the generality of precondition checks, and statistical analysis.

In Section \ref{quantitative-differences}, we analyse the quantitative performance of the two frameworks, demonstrating QuCheck’s improved fault-detection capabilities. QuCheck consistently identified more mutants than QSharpCheck in both the quantum teleportation and superdense coding case studies, with a lower execution time on average.  

\begin{tcolorbox}[colback=blue!5!white,colframe=blue!75!black, enlarge top by=0.2cm, enlarge bottom by=0.2cm, title=Answer to \textbf{RQ3:} How does QuCheck compare to QSharpCheck?]

QuCheck offers a more expressive and extensible testing framework than QSharpCheck, through improvements to property specification, statistical assertions, and a statistical correction for multiple test executions. Quantitatively, QuCheck identified significantly more mutants for quantum teleportation and superdense coding, with a lower average execution time, though it produced one false positive result.

\end{tcolorbox}

\subsubsection{RQ3.1: Qualitative differences}\label{qualitative-differences}

There are a number of qualitative differences that were considered for the evaluation of both frameworks, a summary of the differences can be seen in Table \ref{tab:framework_comparison}

\begin{table}
    \centering
    \begin{tabular}{ p{2.5cm} p{5cm} p{5cm} }
        \hline
        \textbf{Aspect} & \textbf{QuCheck} & \textbf{QSharpCheck} \\
        \hline
        Programming Language & Python (Qiskit) & Q\# \\
        \hline
        Property \par Specification & In a Python class & In a text file \\
        \hline
        Input Generation & 
        Customisable\par
        Deterministic\par
        Various types (i.e., Oracle circuits, Quantum states, Integers)\par
        Multiple qubit states supported &
        Single qubit states can be chained together \par Basic types (i.e. Boolean)\\
        \hline
        Preconditions & Any boolean function using generated inputs & Limited to $\theta$ and $\phi$ of individual qubits \\
        \hline
        Assertions & 
        AssertEqual\par
        AssertProbability\par
        AssertEntangled\par
        AssertDifferent\par
        AssertSeparable\par
        AssertMostFrequent\par \par
        (Multiple basis checks) &
        AssertEqual\par
        AssertProbability\par
        AssertEntangled\par
        AssertTransformed\par
        AssertEqualClassicalBits\par
        AssertTeleported\par \par
        (Computational basis only) \\
        \hline
        Statistical \par Analysis & Statistical correction applied & No statistical correction \\
        \hline
    \end{tabular}
    \caption{Comparison of QuCheck and QSharpCheck testing frameworks.}
    \label{tab:framework_comparison}
\end{table}

\begin{itemize}
    \item \textbf{Programming Language:} The choice of Python and Qiskit for QuCheck was informed by our previous research \cite{PontolilloM22}, which highlighted Qiskit's prevalence in quantum programming repositories. QuCheck builds on the features provided by QSharpCheck, emphasising extensibility to facilitate the creation of custom input generators and assertions tailored to the properties being tested. Additionally, Python's robust package ecosystem further supports this extensibility, enabling seamless integration of external libraries for specialised testing scenarios.
    \item \textbf{Property Specification:} Properties in QSharpCheck are defined in a separate text file, whereas in QuCheck, property specification is done within a python class. This approach enables a higher degree of flexibility when implementing properties: custom input generators can be easily defined and inserted, specialised functions can be used to verify preconditions, and the operation function allows for the definition of more advanced tasks beyond just initialising the circuit.
    \item \textbf{Input Generation:} QSharpCheck's input generation is limited to chaining together single qubit quantum states. QuCheck expands on this by providing a set of predefined deterministic input generators, and an interface to develop new generators. This enables the generation of a wide variety of input types (i.e. oracle circuits), which facilitate the conversion of conceptual properties into property-based tests that can be executed. This capability was pivotal in our case study on Grover's algorithm, where random Grover's oracles that mark a customisable range of states were included as test inputs.
    \item \textbf{Preconditions} In QSharpCheck, preconditions can be used to ensure the randomly generated quantum states fall within a specified range for $\theta$ and $\phi$. Preconditions in QuCheck are a boolean function that receives the randomly generated inputs. If valid inputs cannot be generated within a set number of attempts, QuCheck times out and returns a failure for the property.  
    \item \textbf{Assertions:} QuCheck offers unique assertions such as AssertMostFrequent, AssertDifferent and AssertSeparable. While it does not include AssertTeleported, or AssertEqualClassicalBits, similar results can be achieved through the application of AssertEqual, as exemplified in Listing \ref{lst:Property_example}. Additionally, QuCheck's assertions consider different measurement bases; for AssertEqual, the option to check the X, Y, and Z bases is available, whereas QSharpCheck only measures the computational (Z) basis. Similar to input generation, QuCheck provides an interface for creating custom assertions, which can exploit the extensive Python ecosystem.
    \item \textbf{Statistical analysis:} QSharpCheck performs a fixed number of experiments with a specified number of shots for each test case, and applies a statistical test using the mean of the outcomes. QuCheck conducts one experiment containing all measurement shots and applies the Holm-Bonferroni correction across all statistical tests.
\end{itemize}

\begin{tcolorbox}[colback=blue!5!white,colframe=blue!75!black, enlarge top by=0.2cm, enlarge bottom by=0.2cm, title=Answer to \textbf{RQ3.1:} What are the qualitative differences between QuCheck and QSharpCheck?]
separately. 

QuCheck offers greater expressivity and flexibility than QSharpCheck by defining properties directly in Python files, bypassing QSharpCheck's plain-text parsing limitations. This enables more customisable input generation, precondition checks, and complex operations (e.g., metamorphic properties) before assertions. Designed for extensibility, QuCheck simplifies the addition of new input generators and custom assertions.

QuCheck's predefined assertions are more distinct and versatile, supporting multiple qubits and measurement bases. It expands input generation beyond QSharpCheck's single-qubit states to include entangled multi-qubit states and specialised inputs like Grover's oracles. Additionally, QuCheck can also execute multiple properties in a single experiment with statistical corrections for multiple testing.

\end{tcolorbox}

\subsubsection{RQ3.2: Quantitative differences}\label{quantitative-differences} 

To quantitatively compare the differences of QuCheck and QSharpCheck, we conducted the two case studies that were used for the original QSharpCheck paper. Superdense Coding and Quantum Teleportation. 

\begin{figure}[H]
    \centering
    \includegraphics[width=1\linewidth]{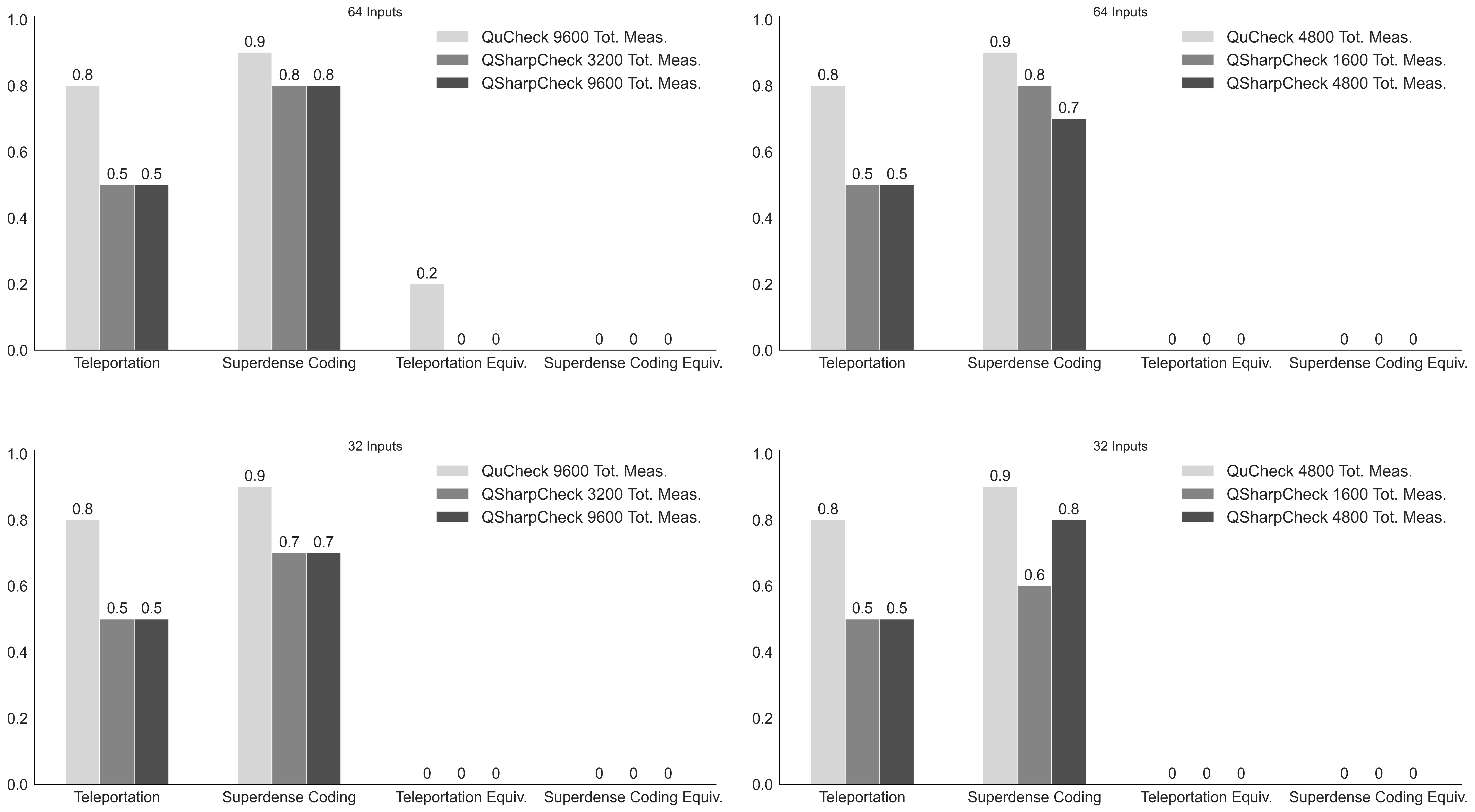}
    \caption{Mutation score comparison between QuCheck and QSharpCheck for Quantum Teleportation, and Superdense Coding.}
    \Description[Four graphs, with the top four input configurations of inputs, comparing the mutation scores for quantum teleportation and superdense coding, for the QMutPy mutants, QuCheck identified more mutants than QSharpCheck. One false positive occurred for the QuCheck set.]{Four graphs, with the top four input configurations of inputs, comparing the mutation scores for quantum teleportation and superdense coding, for the QMutPy mutants, QuCheck identified more mutants than QSharpCheck. One false positive occurred for the QuCheck set.}
    \label{fig:quantitative-comparison}
\end{figure}

The property used for quantum teleportation was the same in both frameworks, with the difference being that our approach verified the output state across three bases. For superdense coding, our approach differed by verifying that the output qubits were in the $|0\rangle$ and $|1\rangle$ states, rather than checking their binary output. This approach enables the detection of phase-related mutations by verifying across three bases. 

The impact of these differences can be seen across all configurations, with the greatest effect on quantum teleportation, where QuCheck always successfully detected three additional mutants out of the ten tested from the QMutPy set (Fig. \ref{fig:quantitative-comparison}). Overall, when averaging across all algorithms, and configurations that performed the same number of \textit{total measurements}, QuCheck achieved an average mutation score 0.85 for the QMutPy mutants, compared to 0.63 with QSharpCheck. However, QuCheck yielded one false positive result within the equivalent mutant set for the quantum teleportation algorithm.

\begin{figure}[H]
    \centering
    \includegraphics[width=1\linewidth]{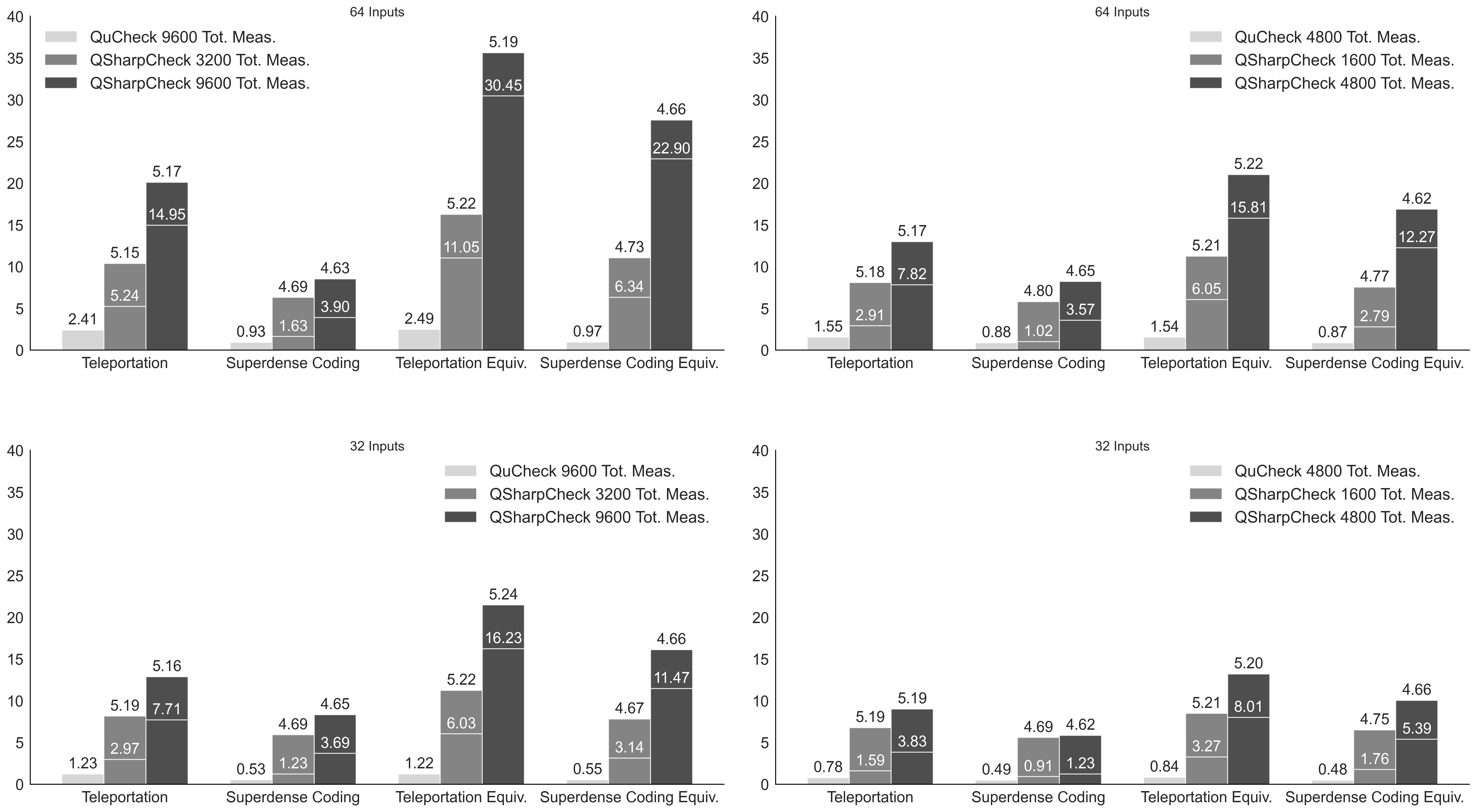}
    \caption{Average total runtime (s) comparison between QuCheck and QSharpCheck for Quantum Teleportation, and Superdense Coding. The lower bars represent execution time, and upper bars represent compilation time.}
    \Description[Four graphs, with the top four input configurations of inputs, comparing the total runtime for quantum teleportation and superdense coding. QuCheck bars portray a lower total runtime across all configurations.]{Four graphs, with the top four input configurations of inputs, comparing the total runtime for quantum teleportation and superdense coding. QuCheck bars portray a lower total runtime across all configurations.}
    \label{fig:quantitative-comparison-time}
\end{figure}

Figure \ref{fig:quantitative-comparison-time} presents the average total runtime for each configuration, with lower bars indicating execution time and upper bars showing compilation time. For \textit{QMutPy mutants}, the median \textit{execution times} for 9600 and 4800 total measurements in QuCheck were 1.09 and 0.83 seconds, compared to 0.56 and 0.41 seconds in QSharpCheck. The standard deviations for these measurements were 0.71 and 0.40 seconds for QuCheck, and 10.33 and 5.50 seconds for QSharpCheck. The low medians but high standard deviations in QSharpCheck are due to execution halting upon the detection of a failure, unlike QuCheck (Sec. \ref{label:limitations}). Despite this, QSharpCheck's execution time was, on average, 81.1\% longer than QuCheck's for the same number of measurements.

For \textit{Equivalent mutants}, no failures were detected in QSharpCheck, so this trend did not occur. Median execution times for 9600 and 4800 measurements in QuCheck were 1.12 and 0.83 seconds, while QSharpCheck recorded 16.93 and 10.02 seconds. The standard deviations were 0.75 and 0.40 seconds for QuCheck, and 7.65 and 4.11 seconds for QSharpCheck. For this set of mutants, QuCheck demonstrates an even greater performance advantage, requiring 92.7\% less time with the same measurement shots.

\begin{tcolorbox}[colback=blue!5!white,colframe=blue!75!black, enlarge top by=0.2cm, enlarge bottom by=0.2cm, title=Answer to \textbf{RQ3.2:} How does QuCheck compare to QSharpCheck in identification of faults in quantum programs?]

QuCheck demonstrated a greater ability to identify faults in QMutPy mutants across the four most thorough test configurations, likely due to its measurement of multiple bases in assertions. It identified 36.0\% more faults with a 81.1\% decrease in average execution time compared to QSharpCheck under measurement parity. There was one instance of a false positive in the equivalent mutant set for QuCheck, which was accompanied by a 92.7\% decrease in execution time.

\end{tcolorbox}

\section{Threats to Validity}\label{threats}

The quantitative analysis between QuCheck and QSharpCheck required the translation of our mutants into Q\#. The translation was partly automated through the use of the Qiskit API and QConvert, this process returned some programs which were not syntactically correct, requiring manual translation that is prone to human error.  

Furthermore, the synthetically generated mutants generated using QMutPy may not accurately represent real faults that would be introduced to these programs by developers. Additionally, the pool of subject systems used for our experiments is of modest size, and it is not clear if the results can be generalised to other quantum programs.   

\section{Conclusion}\label{conclusion}

In this paper, we introduced QuCheck, a property-based testing framework for Qiskit quantum programs. We demonstrated its efficiency and ability to detect faults at varying levels of thoroughness of the test suite. Furthermore, we compared QuCheck to QSharpCheck, highlighting how it addresses QSharpCheck's limitations and offers an accessible property-based testing solution for the Qiskit community. 

Our experiments revealed a positive correlation between the number of measurements, inputs, and properties, and the mutation score, with the number of properties having the most significant impact. The latter finding highlights the importance of covering program semantics through the composition of multiple properties. Conversely, for the equivalent mutant set, we observed a negative correlation between the number of measurements and the mutation score, indicating a reduction in false positives as the statistical power of the tests increased. We also demonstrated the effectiveness and efficiency of the most thorough test configurations along with their associated cost in execution time, confirming the feasibility of applying property-based testing to quantum programs. Finally, our results showed that QuCheck improved fault detection while reducing execution time compared to QSharpCheck. 

Future work will focus on evaluating QuCheck’s performance using real faults instead of synthetic mutants. Additional directions include applying property-based testing with a noisy simulator or real quantum hardware and exploring techniques such as classical shadows \cite{Preskill} for more efficient measurements.

\begin{acks}

The authors have been partially supported by the EPSRC project on Verified Simulation for Large Quantum Systems (VSL-Q), grant reference EP/Y005244/1 and the EPSRC project on Robust and Reliable Quantum Computing (RoaRQ), Investigation 009, grant reference EP/W032635/1. Also, King's Quantum grants provided by King's College London are gratefully acknowledged. The authors thank George Booth, Veronica Gaspes, anonymous referees, and VSL-Q project members for their constructive feedback. 

\end{acks}

\printbibliography

\end{document}